Chapter 1

## MULTIDIMENSIONAL OR RELATIONAL?
## HOW TO ORGANIZE AN ON-LINE ANALYTICAL
## PROCESSING DATABASE

István Szépkúti

*Nationale-Nederlanden Hungary Insurance Co. Ltd.*
*Andrássy út 9, H-1061 Budapest*
*Hungary*
szepkuti@inf.u-szeged.hu

**Abstract**     In the past few years, the number of On-line Analytical Processing (OLAP) applications increased quickly. These applications use two significantly different database structures: multidimensional and table-based. One can show that the traditional model of relational databases cannot make difference between these two structures. Another model is necessary to make the most important differences visible.

One of these differences is the speed of the system. It can be proven, under some reasonable conditions, that the multidimensional database organization results in shorter response times. And it is crucial, since a manager may become impatient, if he or she has to wait say more than twenty seconds for the next screen. On the other hand, here as well as in many other cases, we have to pay for the speed with a bigger database size. Why does the size of multidimensional databases grow so quickly? The reason is the sparsity of data: The multidimensional matrix contains many empty cells. Efficient handling of sparse matrices is indispensable in an OLAP application. One way to handle sparsity is to take the structure closer to the table-based one. Thus the database size decreases, while the application gets slower. Therefore, other methods are needed to eliminate the empty cells from the matrix.

This paper deals with the comparison of the two database structures and the limits of their usage. The new results of the paper can be summarized as follows:

- It gives a constructive proof that all relations can be represented in multidimensional arrays.

- It also shows when the multidimensional array representation is quicker than the table-based one.

- The multidimensional representation results in smaller database size under some conditions. One such sufficient condition is proved in the paper.

- A variation of the single count header compression scheme is described with an algorithm, which creates the compressed array from the ordered table without materializing the uncompressed array.
- The speed of the two different database organizations are tested with experiments, as well. The tests are done on benchmark as well as real life data. The experiments support the theoretical results.

**Keywords:** On-line Analytical Processing, OLAP, multidimensional database, physical database organization.

# 1. INTRODUCTION

## 1.1 DEFINITION

The theory of decision support systems has a long history. Moreover, the first multidimensional language, the "A Programming Language" (APL) dates back to 1962 [16]. On the other hand, the expression On-line Analytical Processing (and its acronym: OLAP) appeared only in 1993. Originally, E. F. Codd et al. described OLAP with twelve evaluation rules [2]:

The twelve rules for evaluating OLAP products are:

1. Multi-Dimensional Conceptual View
2. Transparency
3. Accessibility
4. Consistent Reporting Performance
5. Client-Server Architecture
6. Generic Dimensionality
7. Dynamic Sparse Matrix Handling
8. Multi-User Support
9. Unrestricted Cross-Dimensional Operations
10. Intuitive Data Manipulation
11. Flexible Reporting
12. Unlimited Dimensions and Aggregation Levels

This is a quite lengthy description. Moreover, as their name says, they are rather evaluation rules of software packages than a definition. Fortunately, Nigel Pendse defines the same thing much shortly [17]:

**Definition.** On-line Analytical Processing (or OLAP) means Fast Analysis of Shared Multidimensional Information (shortened: FASMI). □

This definition is much better, because it takes the fact into account that the capacity of the short-term memory of people is seven plus or minus two [15].

Even people with the smallest short-term memory will remember these five words.

E. F. Codd's expression, On-line Analytical Processing, is useful when we want to emphasize the differences between On-line Transaction Processing (OLTP) and OLAP. The most relevant deviations are shown in Table 1.1 [21]:

*Table 1.1*   Comparison of On-line Transaction Processing (OLTP) and On-line Analytical Processing (OLAP).

| Properties | OLTP Business Operations | OLAP Business Intelligence |
|---|---|---|
| Transactions | Many small transactions | Few complex queries |
| Data sources | Internal | Internal and external |
| Time periods | Actual | Historical |
| Queries | Foreseeable, recurring | Unforeseeable, ad hoc |
| Activities | Operative, tactical | Exploratory, analytical, strategic |

## 1.2   RESULTS

The new results of this paper can be summarized as follows:

- It gives a constructive proof that all relations can be represented in multidimensional arrays. (See Assertion 1 in Section 2.) It means that the table representation of relations is only one of the possible ones.

- It also shows when the multidimensional array representation is quicker than the table-based one. In addition, it presents assumptions when the opposite is true. (See Table 1.2 - Table 1.8 and the observations in Section 3.)

- The multidimensional representation results in smaller database size under some conditions. One such sufficient condition is proved in the paper (Assertion 2 in Section 4). The condition is sufficient, but not necessary, because there are several array compression methods.

- A variation of the single count header compression scheme is described with an algorithm, which creates the compressed array from the ordered table without materializing the uncompressed array. (See procedure Array Compression in Section 5.) The compression can be done in one pass, given that the table is sorted either physically or (say through a B-tree index) logically.

■ The speed of the two different database organizations are tested with experiments, as well. The tests are done on benchmark as well as real life data. The benchmark is the TPC-D database, whereas the real life data are originated from an existing company. The experiments support the theoretical results (Table 1.10 - Table 1.12 in Section 5).

## 1.3    RELATED WORK

The compression method used in Section 5 is a variation of the single count header compression scheme (SCH) can be found in [4]. The difference between the two methods is that the SCH accumulates the number of empty cells and the number of nonempty cells separately. These accumulated values are stored in a single alternating sequence. The sum of two consecutive values corresponds to a logical position. Thus, we have to look for a given logical position between these sums. In Section 5, instead of storing a sequence of values, we chose to store pairs of logical positions and number of empty cells: $(L_j, V_j)$. Searching can be done directly on the $L_j$ values; we do not have to sum two consecutive values of a sequence. This results in a simpler searching algorithm, when we want to do logical-to-physical position transformation. On the other hand, if one has to determine the physical position from a $(L_j, V_j)$ pair, then he or she has to take the difference $L_j - V_j$. In case of the SCH, this physical position is explicitly stored; it is nothing else than the accumulated number of nonempty cells. Therefore, the implementation of the physical-to-logical position conversion may be simpler with SCH.

The paper of Zhao et al. [24] uses another compression technique. First, the n-dimensional array is divided into small size n-dimensional chunks. Then, the dense chunks (where the density $\rho > 40\%$) are stored without any modification. Sparse chunks are condensed using "chunk-offset compression." The essence of this method is that only the existing data are stored using (offsetInChunk, data) pairs. Within the chunk, the offset is calculated similarly to the one-dimensional index mentioned in Section 5. The most important difference between this and the previously described technique is that not all the sparse cells are removed from the array. In the pessimistic scenario, when all chunks are just slightly denser than 40%, almost 2.5 times more space is needed to store the cell values, because all empty cells are also stored in this case. This may result in up to 2.5 times more disk input/output operation than absolutely necessary, when the chunks are read or written.

The authors of [24] performed extensive experimentation. They compared, among others, the performance of the cube operator on relational (ROLAP) as well as multidimensional (MOLAP) database organization. The cube is an aggregation operator, which generalizes the group-by clause of an SQL statement. They found that their Multi-Way Array method performs much better than the previously published ROLAP algorithms. Moreover, the performance benefits of the Multi-Way Array method are so substantial that in their tests it was faster to load an array from a table, cube the array, then dump the cubed array into tables, than it was to cube the table directly. In [24], the cube operator was examined, whereas in this paper retrieval is analyzed and tested. It is worth to note the similarities, as well. In [24], just like in this paper, the compressed multidimensional array occupied less space than the table representation; and at the same time, the compressed multidimensional array results in faster operation than the table-based physical representation. In the reasoning at the beginning of Section 3, we mention that the retrieval operation is interesting because the more complicated aggregation operation, uses it in bulk. Therefore, faster retrieval may imply faster aggregation. The experimentation results of [24] supports this statement.

Several papers deal with modeling of OLAP databases. Li and Wang [12] considers dimensions as relations. $D_i$ denotes a dimension name; $r_i$ is the corresponding relation. Then the (n- dimensional) cube is defined as a mapping from $\{\{(D_1, t_1), ..., (D_n, t_n)\} \mid \forall\ 1 \leq i \leq n : t_i \in r_i\}$ to the set $\nu$ of scalar values which includes a special null value. (Please note that this latter "cube" is a mapping by definition, whereas the "cube" in the previous paragraph refers to a generalization of group-by. Despite the identical name, the two concepts are completely different.) In Section 4, we use the concept of conjoint dimensions. The conjoint dimension is a relation itself and the other dimensions can be treated as relations, too. Assertion 1 in Section 2 proves the existence of a one-to-one mapping between the elements of the finite relation R and the nonempty cells of the constructed arrays. In [12], Li and Wang defines the cube as a mapping, whereas we, in the proof of Assertion 1, want to construct a one-to-one mapping to find an alternative physical storage method for relations. Instead of defining the cube concept, we want to store the relations in arrays.

Cabibbo and Torlone [1] defines f-tables (where the "f" stands for "function" or "fact"), which are the logical counterpart of multidimensional arrays. Here the f-table instances are functions by definition.

Gyssens and Lakshmanan [6] introduces the n-dimensional table schema and its instances: the tables. A classical relation corresponds to a 0-dimensional table. A conceptual multidimensional database model is developed in that paper, which is orthogonal to its implementation. Despite this fact, Theorem 2.1 in the paper of Gyssens and Lakshmanan is related to Assertion 1 in Section 2. The theorem says that there is a one-to-one correspondence from the class of tables to the class of relations. The basic difference between the two approaches is that Assertion 1 in Section 2 focuses on the physical representation of relations, whereas the theorem of Gyssens and Lakshmanan stays within the conceptual model.

Li and Wang [12] uses the concept of grouping algebra. Cabibbo and Torlone [1] discusses the design of multidimensional query languages. Lehner [11] proposes the nested multidimensional data model. Gyssens and Lakshmanan [6] utilizes the one-to-one correspondence between tables and relations to "import" the classical operators form the relational algebra. Then the algebra is developed further together with an equivalent calculus. All of these topics are beyond the scope of this paper.

The aforementioned articles on modeling do not deal with the physical representation of the database, which is the most important issue in this paper. In Section 3, extensions (assumptions) are made to the Relational Model in order to enable us to model the different physical representations. Then, based on these assumptions, a kind of a cost model is developed for comparison of the multidimensional and the table-based solutions.

Labio et al. [10] discusses the physical database design for data warehouses. The paper studies how to select the sets of supporting views and indices to materialize in order to minimize the down time because of data refreshments. They call this the view index selection (VIS) problem. However, the paper does not deal with the multidimensional physical representation of relations. In case of a multidimensional array, the indexing is either not necessary, if the array is not compressed, or special indices are needed: if for example the SCH compression is used, the header can be treated as an index. Then this header helps to efficiently retrieve a given cell from the compressed array. The time requirement of this retrieval operation is logarithmic in the size of the header. Thus, the VIS problem probably needs a different handling in case of a multidimensional array. The investigation of the multidimensional VIS problem is also beyond the scope of this paper.

## 1.4 PAPER ORGANIZATION

The rest of the paper is organized as follows. Section 2 deals with modeling issues: how to extend the Relational Model in order to visualize the most important differences between the multidimensional and table-based solutions. Section 3 investigates one crucial attribute of the two methods: the speed of the system. Sparsity is the topic of Section 4. The cost model calculations are supported by experimentation in Section 5. Then we draw the conclusion and summarize the results. The appendix and references can be found at the end of the paper.

## 2. MODELING OLAP DATA STRUCTURES

The Relational Model of databases does not tell anything about the storage of data, as it can be seen in the definition below [7].

**Definition.** The relational database is a finite set of time-dependent relations, which are defined on a finite number of domains.    □

One natural way of storing relations is using a table. Each row of the table means an element of the relation. Therefore this table has to have some special properties [7]:

1. The table must not contain two identical rows.

2. There must be a unique primary key.

3. The order of rows is irrelevant.

4. We refer to the columns with their names.

5. The order of columns is irrelevant.

The first three properties immediately follow from the Relational Model itself. The rest two properties make the table handling more convenient, because we can use the names of the columns instead of their serial numbers.

For the time being, let us forget the table representation of relations and consider the following Sales relation.

$$\text{Sales} \subseteq \text{Geography} \times \text{Product} \times \text{Time} \times \text{Volume},$$

where

$\text{Geography} = \{G_1, G_2, G_3, G_4, G_5\}$
$\text{Product} = \{P_1, P_2, P_3, P_4, P_5, P_6, P_7, P_8, P_9, P_{10}, P_{11}, P_{12}, P_{13}, P_{14}, P_{15}, P_{16}, P_{17}, P_{18}, P_{19}, P_{20}\}$
$\text{Time} = \{T_1, T_2, T_3, T_4, T_5, T_6, T_7, T_8, T_9, T_{10}, T_{11}, T_{12}\}$
$\text{Volume} = \{1, 2, ...\}$

When someone wants to analyze sales data, he or she may need a relation very similar to this one. The Geography domain refers to an area or a region, might be something like {North, South, West, East, Center}. In case of an insurance company, the Product domain could be {Life insurance, Health insurance, Property insurance, Auto insurance, Third party insurance, ...}. The Time domain might contain the months of the actual year: {January 1998, February 1998, March 1998, ...}. The Volume may mean the number of policies sold (or put it another way: the number of insurance contracts concluded), taking the insurance company again, as an example. Now, for us, the actual meaning of the domain elements is not important. The only crucial thing here is that the cardinality of every domain is finite.

Suppose that the first three domains form the unique primary key: Geography, Product and Time. Thus there exists a function $f : \text{Geography} \times \text{Product}$ $\times$ Time $\rightarrow$ Volume, such that f $(G_i, P_j, T_k)$ = v, if and only if $(G_i, P_j, T_k, v) \in$ Sales. Based on this observation, besides the table representation of the Sales relation, we can choose the following alternative method. We put the values of the function f into a three-dimensional matrix or array:

The cell with coordinates (i, j, k)

- has value v, if f $(G_i, P_j, T_k)$ is defined and equals v;

- is empty, otherwise.

The word "otherwise" means that f $(G_i, P_j, T_k)$ is undefined, that is for all v $\in$ Volume the four-tuple $(G_i, P_j, T_k, v) \notin$ Sales.

It is easy to see that the following, more general assertion also holds:

**Assertion 1.** For the finite relation R $\subseteq$ $D_1 \times ... \times D_n$, we can construct multidimensional arrays such that there exists a one-to-one mapping between the elements of R and the nonempty cells of the arrays.

**Proof.** We are going to construct the arrays.

It is enough to consider the case where each domain is finite. Otherwise, let us define $E_i \subseteq D_i$ as follows:

$E_i = \{e_i \mid e_i \in D_i$ such that there exist an n-tuple in R where $e_i$ can be found in the $i^{th}$ position$\}$

Thus R $\subseteq$ $E_1 \times ... \times E_n$ and all $E_i$ domains of R are finite, because R is finite. Now let us replace $D_i$ with $E_i$ and we get the same relation with finite domains. Let $c_i = |D_i|$ and let us denote the elements of $D_i$ in the following way: $D_i = \{d_i(1), ..., d_i(c_i)\}$. Since the order of columns (domains) is irrelevant in the Relational Model, without loss the generality, we can assume that the unique primary key of R consists of $D_1, ..., D_k$ ($1 \le k \le n$). Let us differentiate three cases based on the value of k.

*Case 1.1:* k = n

In this case, all domains can be found in the unique primary key. Therefore, let us construct a k-dimensional (that is n-dimensional) array as follows:

The cell with coordinates $(i_1, ..., i_n)$

- has value 1, if $(d_1(i_1), ..., d_n(i_n)) \in$ R;

- is empty, otherwise.

*Case 1.2:* k = n - 1

Here we have domain $D_n$, which is not part of the primary key. Let us define a function f : $D_1 \times ... \times D_k \rightarrow D_n$ such that f $(d_1(i_1), ..., d_k(i_k)) = d_n(i_n)$, if and only if $(d_1(i_1), ..., d_k(i_k), d_n(i_n)) \in$ R. Then a k-dimensional array can be created in the following way:

The cell with coordinates $(i_1, ..., i_k)$

- ■ has value $f(d_1(i_1), ..., d_k(i_k))$, if it is defined;
- ■ is empty, otherwise.

*Case 1.3:* $k \leq n - 2$

This case is similar to the previous one. The difference is that we can create n - k functions instead of one:

$$f_1 : D_1 \times ... \times D_k \rightarrow D_{k+1}$$
$$...$$
$$f_{n-k} : D_1 \times ... \times D_k \rightarrow D_n$$

In addition, these functions are defined such that $(d_1(i_1), ..., d_k(i_k), d_{k+1}(i_{k+1}), ..., d_n(i_n)) \in R$, if and only if

$$f_1(d_1(i_1), ..., d_k(i_k)) = d_{k+1}(i_{k+1})$$
$$...$$
$$f_{n-k}(d_1(i_1), ..., d_k(i_k)) = d_n(i_n)$$

Hence the number of constructed k-dimensional arrays is also n - k. In the $j^{th}$ array $(1 \leq j \leq n - k)$ the cell with coordinates $(i_1, ..., i_k)$

- ■ has value $f_j(d_1(i_1), ..., d_k(i_k))$, if it is defined;
- ■ is empty, otherwise.

From the construction above, it follows that there exists a one-to-one mapping (a bijective function) between the elements of R and the nonempty cells of the constructed arrays. ■

**Remark.** Let us revisit *Case 1.1* of **Assertion 1** and suppose that R is binary. The thus defined 2-dimensional array is similar to the usual matrix representation of binary relations (see for example [18]). That is why we can say that the definition in *Case 1.1* generalizes the matrix representation of relations.

**Definition.** We are going to call the arrays constructed in **Assertion 1** the multidimensional array representation of relation R. □

**Corollary.** The table representation is only one of the possible physical representations of relations. In some cases, the multidimensional array representation may be as natural as (or even more natural than) the table representation. From the Relational Model point of view it is all the same, because it does not deal with the actual physical representation. So the correct question is "Multidimensional or Table-based?" instead of the one can be found in the title of this article.

## 3.    SPEED

In order to compare the performance of the two significantly different storage methods, we have to model the actual physical representations of the multidimensional and table-based solutions. Throughout this section we are going to investigate only one simple operation: retrieval of one atomic piece of information (either a row of a table or a cell of a multidimensional array). Why are we interested in such a simple operation at all? There are more reasons for this:

1. In many applications, the most important operation is inquiry of data. Matrices of atomic pieces of information are shown on the screen in order to analyze them. Updating the stored information is not typical of these systems. These informational (or decision support) databases are refreshed using batch processing at night. (See the book of Inmon [9].)

2. In some other applications, aggregation and consolidation of data are also commonly used operations (for example in case of budgeting or consolidation of financial figures). These operations are more complicated than retrieval, but in order to do the aggregation and consolidation, we have to inquire the underlying information first. Thus we can say that retrieval is heavily used in case of these more complex operations, as well.

3. Last, but not least, retrieval is simple enough to keep the model clear and mathematically tractable.

To go on with the comparison, we have to make a few assumptions about the physical representations of the relations.

(i) The table and the multidimensional array are stored on the hard disk of the computer.

(ii) In the table-based version, the rows are ordered by the unique primary key.

(iii) In the multidimensional array case, the address of the cell can be calculated from the indices with a formula without accessing the hard disk.

(iv) P denotes the time necessary to position to a given address on the hard disk and to read one atomic piece of information from there.

(v) Multiplication takes time M.

(vi) The number of rows in the table is r.

(vii) In the multidimensional array, the number of dimensions is denoted by k.

The rows of the table are sorted. Hence we can use say a simple binary search to find a given row of the table. The expected number of steps during the binary search is about $log_2$ r - 1. The proof can be found for example in [13]. In every step, we have to read a row from the hard disk and compare its key with the sought one. Comparison is less expensive than positioning and reading. Therefore, we are going to estimate the time necessary to find a row in the table as follows:

$$(log_2 r - 1) \cdot P \qquad (1.1)$$

On the other hand, we can use a formula to calculate the address of a cell in the multidimensional array. Let us consider the formula

$$((...((i_k - 1)c_{k-1} + i_{k-1} - 1)...)c_2 + i_2 - 1)c_1 + i_1 \qquad (1.2)$$

Here $i_j$ means the index of the $j^{th}$ dimension ($1 \leq j \leq$ k and $1 \leq i_j \leq c_j$). Many programming languages use this formula to store k-dimensional arrays in the memory. In our case, we are going to store the k-dimensional array in a file on the hard disk. Then this formula can be used to determine the file position, where the cell is stored.

For instance, let k = 3, $c_1$ = 4, $c_2$ = 3, $c_3$ = 2. In this simple three-dimensional case, it is not difficult to name the dimensions. The first dimension contains the columns. The second includes the rows. The third dimension corresponds to the pages. We can store the cells on the hard disk row by row:

| 1 | 2 | 3 | 4 | 5 | 6 | 7 | 8 | 9 | 10 | 11 | 12 | 13 | 14 | 15 | 16 | 17 | 18 | 19 | 20 | 21 | 22 | 23 | 24 |
|---|---|---|---|---|---|---|---|---|----|----|----|----|----|----|----|----|----|----|----|----|----|----|----|
|   |   |   |   |   |   |   |   |   |    |    |    |    |    |    |    |    |    |    |    |    |    |    |    |
| 1. row | | | | 2. row | | | | 3. row | | | | 1. row | | | | 2. row | | | | 3. row | | | |
| 1. page | | | | | | | | | | | | 2. page | | | | | | | | | | | |

It is easy to check that the address of the cell with coordinates ($3^d$ column, $1^{st}$ row, $2^{nd}$ page) can be calculated as follows:

$$((i_3 - 1) \cdot c_2 + i_2 - 1) \cdot c_1 + i_1 = ((2 - 1) \cdot 3 + 1 - 1) \cdot 4 + 3 = 15 \quad (1.3)$$

This formula can calculate the address of the cell and, similarly to the Horner order for polynomials, we have to multiply only (k - 1) times. In addition, we have to position and read from the hard disk only once. Positioning-and-reading and multiplication are the most costly operations of this case. That is why we are going to estimate the time it takes to retrieve an atomic cell from the multidimensional array in the following way [20]:

$$(k-1) \cdot M + P \qquad (1.4)$$

In order to make the relative performance of the two physical representations visible, let us take the quotient of the two estimations:

$$Q = \frac{(log_2 r - 1) \cdot P}{(k-1) \cdot M + P} \qquad (1.5)$$

This formula depends on four parameters: r, k, P and M. If we want to give values to every parameter independently from each other, then we have to put the values of the formula into a four-dimensional array. Now, you can see that even a simple problem of computer science needs a multidimensional tool to do the analysis! Fortunately, we can simplify the quotient a bit:

$$Q = \frac{(log_2 r - 1) \cdot P}{(k-1) \cdot M + P} = \frac{log_2 r - 1}{(k-1) \cdot \frac{M}{P} + 1} = \frac{log_2 r - 1}{\frac{k-1}{\frac{P}{M}} + 1} = \frac{log_2 r - 1}{\frac{k-1}{p} + 1} \qquad (1.6)$$

This latter quotient depends only on three parameters: r, k and $p := \frac{P}{M}$. We are not interested in the individual values of P (positioning-and-reading) and M (multiplication). They are quite hardware and configuration dependent. We get more general results if we watch only their ratio p. In [20], we obtained that this ratio is approximately 1500 for a given personal computer. Therefore, in Table 1.2 - Table 1.7, we will calculate with p = 1, 10, 100, 500, 1000, 1500. The number of rows (r) will correspond to $10^3$, $10^4$, $10^5$, $10^6$, $10^7$. The number of dimensions (k) will refer to 5, 10, 15, 20, 25. Inside the tables, one can see the quotient Q defined above.

*Table 1.2*   p = 1

| | | | $k$ | | |
|---:|---:|---:|---:|---:|---:|
| $r$ | 5 | 10 | 15 | 20 | 25 |
| 1,000 | 1.79 | 0.90 | 0.60 | 0.45 | 0.36 |
| 10,000 | 2.46 | 1.23 | 0.82 | 0.61 | 0.49 |
| 100,000 | 3.12 | 1.56 | 1.04 | 0.78 | 0.62 |
| 1,000,000 | 3.79 | 1.89 | 1.26 | 0.95 | 0.76 |
| 10,000,000 | 4.45 | 2.23 | 1.48 | 1.11 | 0.89 |

Based on the six tables, we can summarize our observations as follows:

*Table 1.3*   p = 10

| r | | | k | | |
|---|---|---|---|---|---|
| | *5* | *10* | *15* | *20* | *25* |
| 1,000 | 6.40 | 4.72 | 3.74 | 3.09 | 2.64 |
| 10,000 | 8.78 | 6.47 | 5.12 | 4.24 | 3.61 |
| 100,000 | 11.15 | 8.22 | 6.50 | 5.38 | 4.59 |
| 1,000,000 | 13.52 | 9.96 | 7.89 | 6.53 | 5.57 |
| 10,000,000 | 15.90 | 11.71 | 9.27 | 7.67 | 6.55 |

*Table 1.4*   p = 100

| r | | | k | | |
|---|---|---|---|---|---|
| | *5* | *10* | *15* | *20* | *25* |
| 1,000 | 8.62 | 8.23 | 7.86 | 7.53 | 7.23 |
| 10,000 | 11.82 | 11.27 | 10.78 | 10.33 | 9.91 |
| 100,000 | 15.01 | 14.32 | 13.69 | 13.12 | 12.59 |
| 1,000,000 | 18.20 | 17.37 | 16.61 | 15.91 | 15.27 |
| 10,000,000 | 21.40 | 20.42 | 19.52 | 18.70 | 17.95 |

*Table 1.5*   p = 500

| r | | | k | | |
|---|---|---|---|---|---|
| | *5* | *10* | *15* | *20* | *25* |
| 1,000 | 8.89 | 8.81 | 8.72 | 8.64 | 8.56 |
| 10,000 | 12.19 | 12.07 | 11.95 | 11.84 | 11.72 |
| 100,000 | 15.49 | 15.33 | 15.18 | 15.04 | 14.89 |
| 1,000,000 | 18.78 | 18.60 | 18.42 | 18.24 | 18.06 |
| 10,000,000 | 22.08 | 21.86 | 21.65 | 21.44 | 21.23 |

- If we fix the ratio p and the number of dimensions k, then the performance advantage of the multidimensional array is increasing, when the cardinality of the relation (or the number of rows in the table) r is increasing.

- If we keep the ratio p and the cardinality of the relation r fixed, then the speed disadvantage of the table-based representation is decreasing, when the number of dimensions k is increasing.

*Table 1.6*   p = 1000

| r | k | | | | |
|---|---|---|---|---|---|
| | *5* | *10* | *15* | *20* | *25* |
| 1,000 | 8.93 | 8.89 | 8.84 | 8.80 | 8.76 |
| 10,000 | 12.24 | 12.18 | 12.12 | 12.06 | 12.00 |
| 100,000 | 15.55 | 15.47 | 15.39 | 15.32 | 15.24 |
| 1,000,000 | 18.86 | 18.76 | 18.67 | 18.58 | 18.49 |
| 10,000,000 | 22.16 | 22.06 | 21.95 | 21.84 | 21.73 |

*Table 1.7*   p = 1500

| r | k | | | | |
|---|---|---|---|---|---|
| | *5* | *10* | *15* | *20* | *25* |
| 1,000 | 8.94 | 8.91 | 8.88 | 8.85 | 8.82 |
| 10,000 | 12.26 | 12.21 | 12.17 | 12.13 | 12.09 |
| 100,000 | 15.57 | 15.52 | 15.47 | 15.41 | 15.36 |
| 1,000,000 | 18.88 | 18.82 | 18.76 | 18.69 | 18.63 |
| 10,000,000 | 22.19 | 22.12 | 22.05 | 21.98 | 21.90 |

- If the ratio $p \geq 500$, then the multidimensional array representation results in 9 - 22 times faster operation than the table representation, over the investigated area, that is when $10^3 \leq r \leq 10^7$ and $5 \leq k \leq 25$.

$$Q \rightarrow log_2 r - 1 \qquad\qquad (p \rightarrow \infty, \text{ r and k are constants})$$

- When positioning-and-reading is as speedy as multiplication, that is the ratio p is equal to one, and the number of dimensions is large (for example k = 25), then the table-based solution is quicker than the multidimensional approach. Multidimensionality is not omnipotent!

$$Q = \frac{log_2 r - 1}{k} \qquad\qquad (p = 1)$$

In my opinion, in the long run, we have to calculate with this last case, as well. The main reason for this is that the price of memory chips is decreasing. The (personal) computers will contain more and more memory. Contrary to our assumption (i), it will be possible to load the entire table or multidimensional array into the memory. The time requirement of the two operations (positioning-and-reading and multiplication) will be comparable with each other. On the other hand, if we keep assumption (i), and store the table or the array on the hard disk, then the difference between the time requirements of these two operations may be even three orders of magnitude.

If the table is stored on the hard disk, then it is more realistic to assume that the rows of the table are accessed through an index. Let us consider a B-tree index with minimal degree t. In the worst case, we will have to read $log_t \frac{r+1}{2}$ pages from the hard disk to determine the record number of the sought row (see the proof for example in [3]). Then we have to read the row itself. Therefore, the definition of quotient Q has to be modified as follows:

$$Q = \frac{(log_t \frac{r+1}{2} + 1) \cdot P}{(k-1) \cdot M + P} = \frac{log_t \frac{r+1}{2} + 1}{\frac{k-1}{p} + 1} \quad (1.7)$$

In the following table, we are going to assume that positioning-and-reading (P) lasts 1500 times longer than multiplication (M), that is $p = \frac{P}{M} = 1500$. In Section 5, the minimal degree t of the B-tree is 82 in one test database and 89 in another one. We are going to consider the latter value. Then the obtained values of Q are put into Table 1.8.

*Table 1.8*   p = 1500, t = 89

| | | | k | | |
|---:|:---:|:---:|:---:|:---:|:---:|
| r | 5 | 10 | 15 | 20 | 25 |
| 1,000 | 2.38 | 2.37 | 2.36 | 2.35 | 2.35 |
| 10,000 | 2.89 | 2.88 | 2.87 | 2.86 | 2.85 |
| 100,000 | 3.40 | 3.39 | 3.38 | 3.37 | 3.36 |
| 1,000,000 | 3.91 | 3.90 | 3.89 | 3.87 | 3.86 |
| 10,000,000 | 4.42 | 4.41 | 4.40 | 4.38 | 4.37 |

The quotient Q inside the table show smaller values than in the previous case, but its behavior is similar to it. Thus, we can make similar observations.

## 4.    SPARSITY

Multidimensional arrays are usually sparse, that is there are empty cells in them. Why is this so? To understand the reason, let us consider our earlier Sales relation again. Suppose that not a single piece of Product P was sold in Geography G during Time period T. This means that, for all v ∈ Volume, the four-tuple (G, P, T, v) ∉ Sales. Moreover, in a real world situation, this is not an exceptional case at all, so we should not treat it like that.

In the table-based physical representation this means that we will not find a row with unique primary key (G, P, T) in the table. Nothing has to be stored.

But if the relation is represented by a multidimensional array, then the cell with coordinates (G, P, T) will be empty. Moreover, we have to reserve space for each empty cell, as well. (For the time being, we do not deal with the possible compression of the multidimensional array.)

Based on the above argument, we may think that the multidimensional representation always results in a bigger database size. However, the situation is slightly more complicated. Now let us take the general case and consider the finite relation $R \subseteq D_1 \times ... \times D_n$, where the domains $D_1$, ..., $D_k$ ($1 \leq k \leq n$) form the unique primary key. The cardinality of domains is denoted by $c_i = |D_i|$, whereas the cardinality of the relation is $r = |R|$. In order to decide whether the table-based or the multidimensional method results in smaller space requirement, let us make the following additional assumptions:

(viii)  The space requirement of one row in the table (say in bytes) is S.

(ix)  The space requirement of $D_{k+1}$, ..., $D_n$ (that is the domains outside the key) within a row is $\delta \cdot S$, where $0 \leq \delta < 1$.

From these assumptions it follows that size ratio of non-key columns and all columns (within a row) is $\delta$. Columns outside the key contain the actual data, which we want to analyze. Therefore, let us call $\delta$ the data ratio. If we want to store the relation in a table, then we need $r \cdot S$ space, because there are r rows and one row occupies S space. In the opposite case, when the relation is stored in a k-dimensional array, the space requirement is $c_1 \cdot ... \cdot c_k \cdot \delta \cdot S$. Now let us divide this latter formula by the former one in order to compare the multidimensional space requirement ($S_m$) with the table-based one ($S_t$):

$$\frac{multidimensional\ space\ requirement}{table - based\ space\ requirement} = \frac{S_m}{S_t} = \frac{c_1...c_k \delta S}{rS} = \frac{c_1...c_k \delta}{r} \tag{1.8}$$

To measure the sparsity - density of the multidimensional array, let us introduce another ratio $0 < \rho \leq 1$, which will denote the density of the array:

$$\rho = \frac{number\ of\ nonempty\ cells}{total\ number\ of\ cells} = \frac{r}{c_1...c_k} \tag{1.9}$$

The definitions of the data ratio ($\delta$) and the density ($\rho$) imply immediately that

$$\frac{S_m}{S_t} = \frac{c_1...c_k \delta}{r} = \frac{\delta}{\rho} \tag{1.10}$$

The connection between the space requirements of the two different physical representations can be stated more precisely as follows:

**Assertion 2.** The multidimensional physical representation results in a smaller database size, if the data ratio ($\delta$) of the table is smaller than the density ($\rho$) of the multidimensional array.

**Proof.** This assertion is a direct consequence of equation (1.10).    ■

This **Assertion 2** is a robust statement in the sense that it holds regardless the fact whether the multidimensional array is compressed somehow (in order to eliminate empty cells) or not. On the other hand, it is not difficult to calculate the data ratio from the data types of the domains $D_1, ..., D_n$, given that we know which domains constitute the unique primary key. It is also easy to determine the density; we just have to count the number of elements of $D_1, ..., D_k$ and R.

Let us consider the situation when Product P is not sold in Geography G at all. (Say the insurance company does not want to sell auto insurance in the central region, because it is too risky and unpredictable.) In this case, the multidimensional array will contain at least twelve empty cells, because for all T ∈ Time, all cells with coordinates (G, P, T) will not contain any actual value. One way to get rid of these surely empty cells is to form a so-called conjoint dimension from the existing elements of Geography × Product:

Conjoint = {(G, P) | (G, P) ∈ Geography × Product and there exist T ∈ Time and v ∈ Volume such that (G, P, T, v) ∈ Sales} = $\pi_{Geography, Product}$(Sales)

Having this Conjoint constructed, we can define a Sales' relation, which is equivalent to the original Sales relation:

Sales' = {((G, P), T, v) | ((G, P), T, v) ∈ Conjoint × Time × Volume such that (G, P, T, v) ∈ Sales}

Sales and Sales' are just the same relation; their table representation is also very similar to each other. However, the multidimensional arrays corresponding to them will be different. The total number of cells in the second case will be less, thus increasing the density of the array.

Taking a slightly more general approach, let us suppose that the finite relation R has a special property: given elements of $D_1 \times ... \times D_h$ ($1 \leq h \leq k$) cannot be found in the corresponding projection of R. Thus, in order to eliminate empty cells form the multidimensional array representation, we can define an equivalent R' relation:

R' = {((d_1, ..., d_h), d_{h+1}, ..., d_n) | ((d_1, ..., d_h), d_{h+1}, ..., d_n) ∈ Conjoint × $D_{h+1} \times ... \times D_n$ such that (d_1, ..., d_h, d_{h+1}, ..., d_n) ∈ R}

where

$$\text{Conjoint} = \pi_{D_1, ..., D_h}(R)$$

Here as well as in the previous case, $\pi$ denotes the projection operation of relations. Now, we are ready to calculate the total number of cells per array in case of the original and the new relations.

Total number of cells in case of R = $N_R$ = $c_1 \cdot ... \cdot c_k$

Total number of cells in case of R' = $N_{R'}$ = |Conjoint| $\cdot c_{h+1} \cdot ... \cdot c_k$

We can characterize the decrease with the following quotient:

$$\frac{N_{R'}}{N_R} = \frac{|Conjoint|c_{h+1}...c_k}{c_1...c_k} = \frac{|Conjoint|}{c_1...c_h} \quad (1.11)$$

So the total number of cells is decreasing but the number of nonempty cells is the same. That is why the density of the multidimensional array is increasing from $\rho$ to $\rho$':

$$\rho' = \frac{c_1...c_h}{|Conjoint|}\rho \quad (1.12)$$

We have to be careful with conjoint dimensions. Consider, for example, the case when h = k, that is all elements of the unique primary key are put into Conjoint. One can see that we could eliminate all empty cells this way and the multidimensional representation became identical with the table-based one! Thus, we have to exclude this extreme case of Conjoint, because it probably degrades the performance.

## 5.    EXPERIMENTS

Several experiments were done to verify the relevancy of the obtained results on the speed of the different database organizations. The data used for testing were originated from two sources:

- The transactional database of a company. In order to keep trade secret, the name of this company is not specified in this paper. To simplify the reference to this enterprise, let us call it insurance company.

- The TPC-D benchmark database [22].

There are several advantages and disadvantages of these two sources. The TPC-D benchmark database can be generated with a freely available program called DBGEN. The size of the output can be given through a parameter (scale factor) to this program. It is easy to compare the results reported in different papers, if the same standard benchmark database is used in them. For example: the TPC-D benchmark database is used in [8, 5], etc. On the other hand, the TPC-D benchmark database contains randomly generated data, which may make it quite different from a real life database. That is why, it still makes sense to examine the performance of the different physical database organizations on the database of the insurance company. These latter results will be less comparable, if comparison is possible at all. Most of the enterprises (just like the insurance company) will not allow anyone to inspect their data because of trade and other secrets.

Using DBGEN, we generated a database of 100 Megabytes (the scale factor was equal to 0.1). The hardware and software used for testing are described in Appendix A. The characteristics of the data used for testing can be found in Table 1.9:

*Table 1.9*  Comparison of the two databases.

|  | Database of the insurance company | TPC-D benchmark database |
|---|---|---|
| Number of dimensions | 5 | 3 |
| Cardinality of the relation | 150,412 | 600,350 |
| Size of table representation | 10,370 KB | 39,014 KB |
|     Table | 2,938 KB | 11,726 KB |
|     B-tree index | 7,432 KB | 27,288 KB |
| Size of array representation | 1,471 KB | 11,912 KB |
|     Compressed array | 294 KB | 2,346 KB |
|     Header | 1,175 KB | 9,374 KB |
|     Dimension values | 2 KB | 192 KB |

In case of the insurance company, two other dimensions were included in the database in addition to the usual three: Geography, Product and Time. Similarly to the solution of [8] and [5], a three-dimensional relation was created from the TPC-D benchmark database with dimensions: Part, Supplier and Customer.

One relation was tested per database. The cardinality of the relation is equal to the number of rows in the table representation, and it equals the number of nonempty cells in the multidimensional (array) representation.

The table representation consists of a table and a B-tree index. The index was used to speed up the access to a given row of the table. The table and the index were stored on the hard disk. That is why we expect that the B-tree search is quicker than the binary search. The minimal degree of the B-tree was 82 in case of the database of the insurance company and 89 in case of the TPC-D benchmark database.

The multidimensional representation includes one compressed array, one header and one file for each dimension to store the dimension values. All empty cells are removed from the array in order to save space. The logical-to-physical position conversion is implemented with the help of the header, which is stored in a separate file. First, the n-dimensional indices are transformed into a one-dimensional index according to the formula can be found in Section 3:

$$((...((i_k - 1)c_{k-1} + i_{k-1} - 1)...)c_2 + i_2 - 1)c_1 + i_1 \qquad (1.13)$$

After this transformation, we get a sequence of empty and nonempty cells:

$$(E*N*)*$$

In the above regular expression, E denotes an empty cell, whereas N denotes a nonempty one. For each E*N* run, the logical position of the last N is determined as well as the total number of empty cells before this position. These $(L_j, V_j)$ pairs of values are ordered by logical position and stored in a table, which is kept in memory. This table is called header.

The compression can be done based on the table representation, without actually creating the sparse array. The following procedure describes this algorithm in pseudo-code:

**procedure** Array Compression;
**begin**
    Initialize dimensions and variables;
    Open the ordered table for input, the compressed array
    and the header for output;
    **while not** Eof(table) **do**
    **begin**
      Save the record number of the table;
      Read a row from the table;
      Write the measure attribute of the row
      into the compressed array;
      i := $((...((i_k - 1)c_{k-1} + i_{k-1} - 1)...)c_2 + i_2 - 1)c_1 + i_1$;
      **if** the current i is greater than the previous i + 1 **then**
      **begin**
        Number of empty cells before the previous i
          := previous i - (record number - 1);
        Write the run ending with the previous i into the header
        as a pair of (previous i, number of empty cells before
        the previous i)
      **end**;
      Previous i := i;
    **end**;
    i := $c_k c_{k-1} ... c_2 c_1$;
    **if** the current i is greater than the previous i + 1 **then**
    **begin**
      Number of empty cells before the previous i

```
        := previous i - total number of records;
    Write the run ending with the previous i into the header
    as a pair of (previous i, number of empty cells before
    the previous i)
  end;
  Number of empty cells before the current i
    := current i - total number of records;
  Write the run ending with the current i into the header
  as a pair of (current i, number of empty cells before
  the current i);
  Close all files
end;
```

This procedure assumes that the one-dimensional index i is calculated such that $1 \leq i \leq c_k c_{k-1} \ldots c_2 c_1$ for all i within the boundaries of the array. In addition, it starts the record numbering, as well, from 1. The table may be ordered either physically or logically (through a B-tree index). When we want to find a given cell in the array, we have to do the following:

1. Transform the n-dimensional indices into a one-dimensional index (the latter will be denoted by i).

2. In the aforementioned table of logical positions and number of empty cells (header), find the $(L_j, V_j)$ pair corresponding to the one-dimensional index. If exact match is not found, then search for the first logical position, which is greater than the one-dimensional index. As a result, the following inequality will hold: $L_{j-1} < i \leq L_j$.

3. Calculate the physical position of the cell from the logical position and determine the value of the desired cell as follows:

   ■ If $L_{j-1} + V_j - V_{j-1} < i$, then the cell is not empty and its value can be retrieved from the compressed array from physical position $i - V_j$;

   ■ Otherwise the cell is empty.

Before compression, the array looks like this:

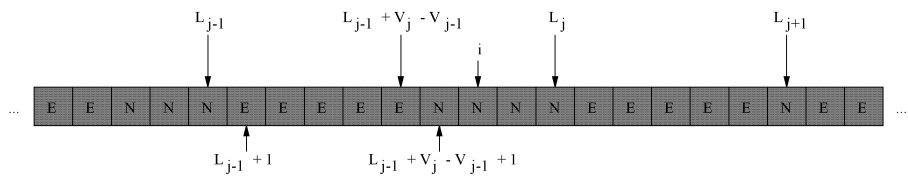

The label at the beginning of the arrow shows the logical position of the cell, which can be found at the end of the arrow.

After compression, the empty cells are entirely removed from the array (the arrows are labeled by the physical position of the cell):

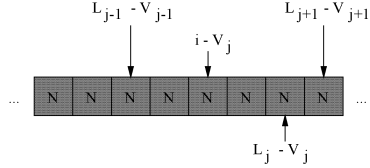

The number of empty cells in the $j^{th}$ run is equal to $V_j - V_{j-1}$, because $V_j$ equals the total number of empty cells up to and including the $j^{th}$ run. Similarly, $V_{j-1}$ is the number of empty cells before the end of the $(j-1)^{th}$ run. That is, from logical position $L_{j-1} + 1$ to logical position $L_{j-1} + V_j - V_{j-1}$, the $j^{th}$ run contains empty cells only. From logical position $L_{j-1} + V_j - V_{j-1} + 1$ to logical position $L_j$, all cells in the $j^{th}$ run are nonempty.

With the use of this compression, we could reduce the storage requirement experienced at the table representation to its 14% in case of the database of the insurance company and to 31% in case of the TPC-D benchmark database. In the former case, the header contained four-byte integer value pairs, whereas in the latter case the header consisted of eight-byte integer value pairs. On the other hand, we had to introduce a search phase for the logical-to-physical position conversion. In both cases, the header was cached into the memory. Thus, we could use binary search to find the necessary logical position.

From the databases, we took random samples with repetitions of the following sizes: 100, 500, 1,000, 5,000, 10,000, 50,000 and 100,000. Uniform distribution was applied: the rows (the nonempty cells) were sampled with equal probabilities. Then the elements of the random samples were sought, one by one, in the table through the B-tree index and in the compressed array with the use of the header. The elapsed time was measured. The time it takes to find the elements of the sample in the table was compared to the time it takes to find them in the compressed array by taking the quotient of the two elapsed times. The comparison gave the result can be found in Table 1.10:

*Table 1.10*   Comparison of the multidimensional and the table representations.

| Sample size | 100,000 | 50,000 | 10,000 | 5,000 | 1,000 | 500 | 100 |
|---|---|---|---|---|---|---|---|
| Insurance co. | 6.22 | 8.41 | 21.50 | 22.06 | 27.56 | 18.92 | 7.82 |
| TPC-D | 51.52 | 35.61 | 15.26 | 7.85 | 2.11 | 1.64 | 1.54 |

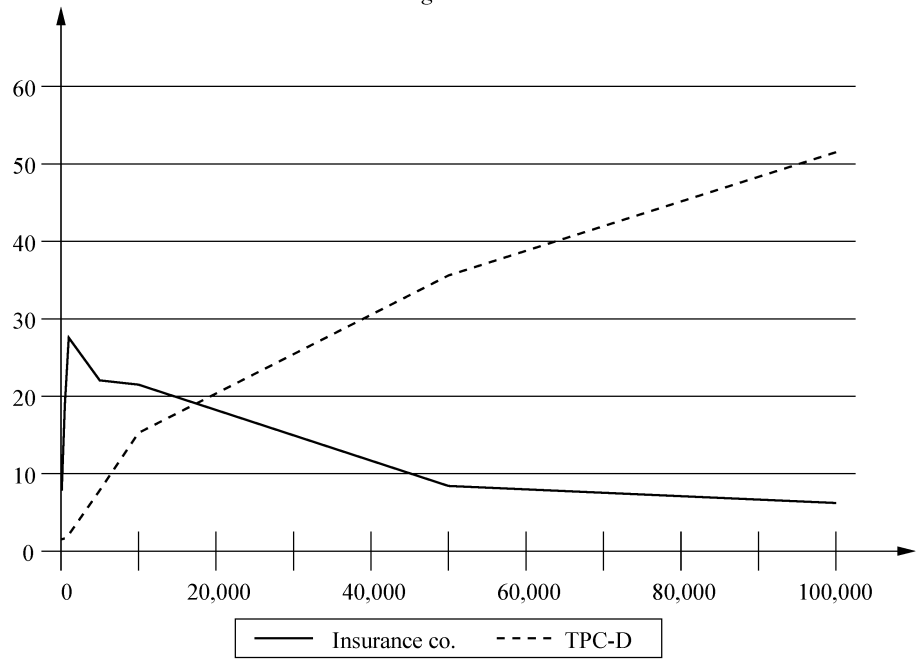

Figure 1.1

The table shows these quotients inside, as a function of the sample size. One can see in the table that the cells could be retrieved 6 - 28 times faster from the multidimensional physical representation than the rows from the table representation, if the database of the insurance company was tested. If the TPC-D benchmark database was sampled, then the multidimensional representation was 1.5 - 52 times quicker than the table-based one. These functions are depicted in Figure 1.1.

The sizes of the two databases are different. Therefore, a given sample size corresponds to different proportion of the databases. For example: a sample size of 100,000 corresponds to the 66.48% of the database of the insurance company, whereas such sample size covers only the 16.66% of the TPC-D benchmark database. The results of the experiments as a function of the sample percentage (sample size divided by the cardinality of the relation) are shown in Table 1.11 and Table 1.12.

If we use the sample percentage as the horizontal axis, then we get the chart in Figure 1.2.

Both charts show that the quotient of elapsed times depends on the size of the sample. This function-like dependence cannot be explained by or model. The disk caching performed by the operating system causes it probably. Big portion of the relatively small multidimensional representation is retained in the disk cache. This may be the reason that sometimes the multidimensional physical representation is even faster than expected.

*Table 1.11* The results as a function of the sample percentage in case of the database of the insurance company.

| Sample % | 66.48 | 33.24 | 6.65 | 3.32 | 0.66 | 0.33 | 0.07 |
|---|---|---|---|---|---|---|---|
| Insurance co. | 6.22 | 8.41 | 21.50 | 22.06 | 27.56 | 18.92 | 7.82 |

*Table 1.12* The results as a function of the sample percentage in case of the TPC-D benchmark database.

| Sample % | 16.66 | 8.33 | 1.67 | 0.83 | 0.17 | 0.08 | 0.02 |
|---|---|---|---|---|---|---|---|
| TPC-D | 51.52 | 35.61 | 15.26 | 7.85 | 2.11 | 1.64 | 1.54 |

*Figure 1.2*

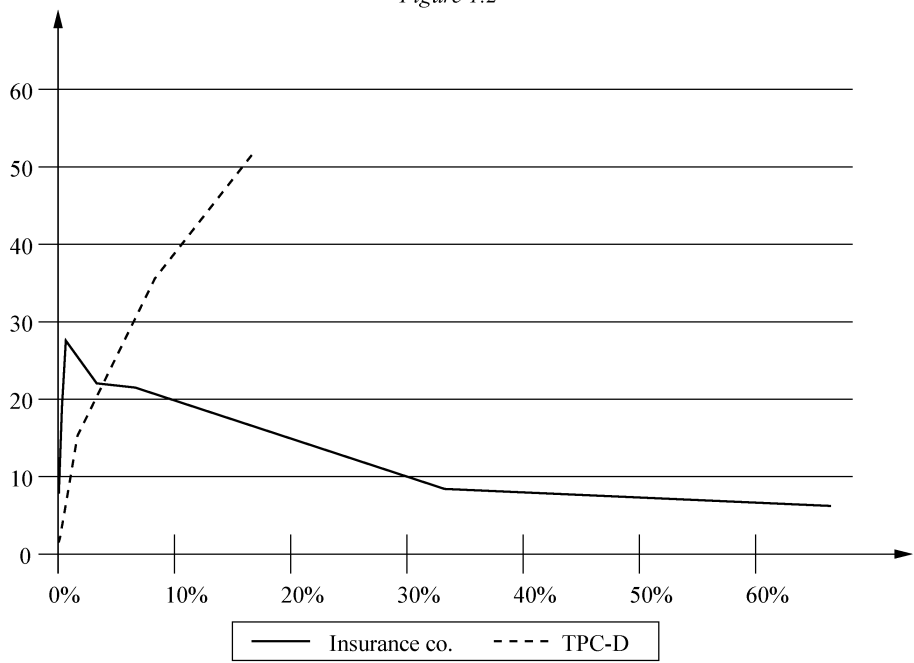

## 6.    CONCLUSION

Several articles, just like the title of this one, raise the question like this: "Multidimensional or Relational?"  But the approach of this question is not correct, because the multidimensional physical representation can perfectly fit into the logical Relational Model.  Agreeing with E. F. Codd [2], we can say that the physical schema and the logical and conceptual schema should not be confused with each other.

Both multidimensional and table-based approaches have their own place within OLAP. In some cases, the table-based approach is better.  In other cases the multidimensional approach is outperforming the table representation. Hence, an OLAP tool has to support multidimensional and table-based as well as hybrid solutions in order to survive in the long run.

The results of this paper can be summarized as follows:

- It shows that the multidimensional physical representation of relations may be as natural as the table-based one.

- It sets up a deliberately simplistic model of the physical representations of On-line Analytical Processing databases.  Within this model, it shows that in many cases the multidimensional solution is speedier than the table-based method.  At the same time, there are cases, when the table-based physical representation is more beneficial than the other one.

- It proves that there are cases, when the multidimensional physical representation results in smaller database size than the table-based physical representation.

- It presents a variation of the single count header compression scheme.

- The results gained from the cost model are supported by the experiments.

## Acknowledgments

I would like to thank János Csirik for his invaluable comments on earlier versions of this paper.  I would also like to thank Tibor Csendes and Zoltán Kincses for thier useful suggestions.

# Appendix

*Table 1.A.1*    The table shows the hardware and software, which were used for testing.

| | |
|---|---|
| Computer | Toshiba Satellite 300CDS |
| Processor | Intel Pentium MMX |
| Processor speed | 166 MHz |
| Memory size | 32 MB |
| Average access time of hard disk | 14 ms |
| File system | FAT32 |
| Page size | 4 KB |
| Operating system | Microsoft Windows 98 4.10.1998 |
| Development environment | Borland Delphi Standard Version 3.0 |
| Programming language | Pascal |

**Remark.** Most of the latest papers referenced here can be downloaded through the [14] bibliography.